\documentclass[11pt,twoside]{article}

\usepackage{graphicx}
\usepackage[misc]{ifsym}
\usepackage{amsfonts}
\usepackage{amssymb,amsbsy}
\usepackage{amsmath,amssymb,amsthm,fancybox,array}
\usepackage{enumerate}
\usepackage[utf8]{inputenc}
\usepackage[T1]{fontenc}
\usepackage{pxfonts}
\usepackage{dsfont}
\usepackage{float}
\usepackage[top=70mm, bottom=30mm, left=30mm, right=40mm]{geometry}

\theoremstyle{plain}
\newtheorem{theorem}{Theorem}[section]

\theoremstyle{definition}

\theoremstyle{remark}

\newtheorem{theproof}{\textbf Proof.}

\numberwithin{equation}{section}

\begin{document}
\setcounter{page}{1}
\thispagestyle{empty}

\markboth{~~\hrulefill~~ \small The Author(s)} 
{\small Inverse Weibull Distribution  ~~\hrulefill~~ }

\topmargin=0mm
\vspace{2cm}
{\centering
{\Large Estimation of the inverse Weibull distribution under type-I hybrid censoring}\\[0.7cm]}
{\bf \noindent Mohammad Kazemi $^1$, Mina Azizpoor $^2$}\\ \\ 
{\small  $^1$ Department of Statistics, School of Mathematical Sciences, Shahrood University of Technology, Shahrood, Iran \\
$^2$ Department of Statistics, Faculty of Mathematical Sciences, University of Mazandaran, Babolsar, Iran}

\renewcommand{\thefootnote}{}
\footnotetext{\hskip-0.6cm Corresponding Author(\Letter)(email), M.Kazemie64@yahoo.com}

\vspace{0.7cm}
{\noindent \bf Abstract.}
 The hybrid censoring is a mixture of Type I and Type II censoring schemes. This paper presents the statistical inferences of the Inverse Weibull distribution when the data are Type-I hybrid censored. First we consider the maximum likelihood estimators of the unknown parameters. It is observed that the maximum likelihood estimators can not be obtained in closed form. We further obtain the Bayes estimators and the corresponding highest posterior density credible intervals of the unknown parameters under the assumption of independent gamma priors using the importance sampling procedure. We also compute the approximate Bayes estimators using Lindley's approximation technique. We have performed a simulation study and a real data analysis in order to compare the proposed Bayes estimators with the maximum likelihood estimators.  \\[2mm]

\renewcommand{\thefootnote}{\arabic{footnote}}
\setcounter{footnote}{0}

\vspace*{0.3cm}
\section{Introduction}

 Type-I and Type-II are the two most popular censoring schemes which are in use for any life testing experiment. Two mixtures of Type-I and Type-II censoring schemes are known as hybrid censoring schemes. If the experiment terminates as soon as either the $R$-th failure or the pre-specified censoring time $T$ occurs, type-I hybrid censoring scheme has been performed. In type-II hybrid censoring scheme, the experiment terminates when the latter of the $R$-th failure and the censoring time $T$ occurs. Denote the $i$-th order statistic from a random sample of size $n$ by $X_{i:n}$. Thus, in type-I hybrid censoring scheme, one observes $X_{1:n},\cdots ,X_{r:n}$ when $X_{r:n}\leq \min\{X_{R:n},T\}$ and $X_{r+1:n}> \min\{X_{R:n},T\}$. Under this scheme, the experiment may be terminated too early resulting in very few failures. Under type-II hybrid censoring scheme, the experiment terminates when
$X_{1:n},\cdots ,X_{r:n}$ are observed for which $X_{r:n}\leq \max\{X_{R:n},T\}$ and $X_{r+1:n}> \max\{X_{R:n},T\}$. In both hybrid censoring schemes, the failure number $R$ and censoring time $T$ are pre-fixed.\

Epstein (1954) first introduced the hybrid censoring scheme and analyzed the data under the assumption of exponential lifetime distribution of the experimental units. An extensive literature exists for hybrid censoring under classical and Bayesian framework and the overview presented below describe
some of the work done on this topic. Gupta and Kundu (1998) obtained confidence and credible intervals for an one-parameter exponential distribution. Kundu (2007) obtained the MLE's, the approximate MLE's and Bayes estimates of shape and scale parameters of a Weibull distribution. Kundu and Pradhan (2009) analyzed a generalized exponential distribution in presence of hybrid censoring. Balakrishnan and Shafay (2012) developed a general method for obtaining Bayes prediction intervals of future observable based on an observed Type-I hybrid censored data. Rastogi and Tripathi (2013) derived maximum likelihood and Bayes estimates of the unknown model parameters of a Burr XII distribution. Singh and Tripathi (2015) studied a two-parameter lognormal distribution using
hybrid censored samples and derived various point and interval estimates of unknown lognormal parameters from classical and Bayesian viewpoint. Tripathi and Rastogi (2015) considered point and interval estimation of the unknown parameters of a generalized inverted exponential distribution and obtained various classical and Bayes estimates based on hybrid censored samples. Hyun, Lee and Robert (2016) analyzed a two-parameter log-logistic distribution based on type I and type II hybrid censored data.

In this paper, we provide point and interval estimators for the unknown parameters of an inverse Weibull (IW) distribution based on type-I hybrid censored samples. The probability density function (PDF) of an IW distribution is
\begin{equation}\label{q1}
f_X(x;\alpha,\lambda)=\frac{\alpha}{\theta^{\alpha}} x^{-(\alpha+1)}e^{-(\theta x)^{-\alpha}},~~~~~~~~x>0,
\end{equation}
and the corresponding cumulative distribution function (CDF) is given by
\begin{equation}\label{q2}
F_X(x;\alpha,\lambda)=e^{-(\theta x)^{-\alpha}},~~~~~~~x>0,
\end{equation}
where $\alpha>0$ and $ \theta >0 $  are the shape and scale parameters respectively. As in the Weibull distribution, the shape parameter $\alpha$ governs the shape of the PDF, the hazard function and the general properties of the IW distribution. When  $\alpha = 1$ and $\alpha = 2 $, the IW distribution reduce to the inverse exponential and inverse Rayleigh distributions respectively. \

Extensive work has been done on the IW distribution. Kundu and Howlader (2010) considered the Bayesian inference and prediction problems of the
IW distribution based on Type-II censored data. Singh et al. (2013) proposed a Bayesian procedure for the estimation of the parameters of IW distribution under Type-II hybrid censoring scheme. Ateya (2015) considered point and interval estimation of the unknown parameters of a IW distribution based on Balakrishnan’s unified hybrid censoring scheme. We consider the inference for IW distribution under type-I hybrid censoring scheme.  \

The rest of the paper is organized as follows. In Section 2, we discuss the maximum likelihood estimation of the scale and shape parameters of the IW distribution. The asymptotic confidence bounds are provided in section 3. Bayesian analyses are presented in Section 4. In Section 5, we conduct a simulation study to compare the performance of proposed methods and then analyzed a real data set for illustrative purpose in Section 6. Finally we conclude the paper in section 7.

\section{Maximum Likelihood Estimation}

In this section we provide the maximum likelihood estimators (MLEs) of the unknown parameters. We re-parametrize the model as follows $ \lambda=\frac{1}{\theta^{\alpha}} $. Suppose $ n $ identical units are put on life test. Then under type-I hybrid censoring scheme, we observe only the first $ r $ failure times, say $ t_1,t_2, \ldots, t_r $. Under the assumptions that the lifetime distribution of the items are independent and identically distributed (i.i.d.) IW random variable, the likelihood function for the type-I hybrid censored data without the multiplicative constant can be written as

\begin{equation}\label{q3}
L(\alpha,\lambda\mid data)=\alpha^{r}\lambda^{r}e^{-\lambda \sum_{i=1}^r x_i^\alpha}\prod_{i=1}^r x_i^{\alpha+1} (1-e^{-\lambda u^{-\alpha}})^{n-r},
\end{equation}
where $ x_i=\frac{1}{t_{(i)}} $, $u=min(t_{(R)}, T)$ and $ r $ denotes the number of units that would fail before the time $ u $. Taking the logarithm of (\ref{q3}), we obtain
\begin{equation}\label{q4}
l(\alpha,\lambda\mid data)=r\ln (\alpha \lambda) -\lambda\sum_{i=1}^r {x_{i}^{\alpha}}+(\alpha +1)  \sum_{i=1}^r {\ln {x_{i}}}+(n-r) \ln(1-e^{-\lambda u^{-\alpha}}).
\end{equation}

Taking derivatives with respect to $ \alpha $ and $\lambda$ of (\ref{q4}), and equality to zero, we obtain
\begin{align}\label{q5}
\frac{\partial \ln L}{\partial \alpha}&=\frac{r}{\alpha} -\lambda \sum_{i=1}^r {x_{i}^{\alpha} \ln x_{i}}
+\sum_{i=1}^r \ln x_{i}+(n-r)\frac{\lambda u^{-\alpha} \ln u e^{-\lambda u^{-\alpha}}}{1 -e^{-\lambda u^{-\alpha}}},\nonumber\\
\frac{\partial \ln L}{\partial \lambda}&=\frac{r}{\lambda} - \sum_{i=1}^r {x_{i}^{\alpha}}
+(n-r)\frac{ u^{-\alpha} e^{-\lambda u^{-\alpha}}}{1 -e^{-\lambda u^{-\alpha}}}.
\end{align}

It is clear that the normal equations do not have explicit solutions. We need some numerical techniques to solve the simultaneous equations.

\section{Asymptotic Confidence Bounds}
\paragraph{ }

Since the MLEs of the unknown parameters $ \alpha, \lambda $ can not be obtained in closed forms, it is not easy to derive the exact distributions of the MLEs.  Therefore, the exact confidence intervals for the unknown parameters is difficult to obtain. In this section, we compute the observed Fisher information based
on the likelihood equations. These will enable us to develop pivotal quantities based on the limiting normal distribution and then construct asymptotic confidence intervals.\

From the log-likelihood function in (\ref{q4}) we obtain the observed Fisher information as

\begin{equation*}
\frac{\partial ^{2} l(\alpha,\lambda)}{\partial \alpha^2 }=- \frac{r}{\alpha^2}-\lambda\sum_{i=1}^r {x_{i}^{\alpha}( \ln x_{i})^2}+\frac{(n-r) \lambda u^{-\alpha}(\ln u)^2 e^{-\lambda u^{-\alpha}}(1-\lambda u^{-\alpha})}{1-e^{-\lambda u^{-\alpha}}} \nonumber
\end{equation*}
\begin{align*}
&-\frac{(n-r) \lambda^{2} u^{-2\alpha} \ln u^{2} e^{-2\lambda u^{- \alpha}}}{{(1-e^{-\lambda u^{-\alpha}})}^2},\nonumber\\
\frac{\partial ^{2} l(\alpha,\lambda)}{\partial \alpha \partial \lambda }&=-\sum_{i=1}^r {x_{i}^{\alpha} \ln x_{i}}
-\frac{(n-r) u^{-\alpha}  \ln (u) e^{-\lambda u^{-\alpha}}(1-\lambda u^{-\alpha})}{1-e^{-\lambda u^{-\alpha}}}\nonumber\\
&+\frac{(n-r) u^{-2\alpha} \ln (u)  \lambda e^{-2\lambda u^{-\alpha}}}{(1-e^{-\lambda u^{-\alpha}})^2 }, \nonumber\\
\frac{\partial ^{2} l(\alpha,\lambda)}{\partial \lambda^2 }&=- \frac{r}{\lambda^2}-\frac{(n-r) u^{-2\alpha} e^{-\lambda u^{-\alpha}}}{1-e^{-\lambda u^{-\alpha}}}-\frac{(n-r) u^{-2\alpha} e^{-2\lambda u^{- \alpha}}}{{(1-e^{-\lambda u^{-\alpha}})}^2}. \nonumber
\end{align*}

The observed Fisher information matrix can be inverted to obtain the asymptotic
variance-covariance matrix of the MLEs as\\

\begin{equation}\label{qq1}
\mathbf{I^{-1}} =\left\{- \left(
\begin{array}{cc}
\frac{\partial ^{2} l(\alpha,\lambda)}{\partial \alpha^2 }\mid_{\hat{\alpha},\hat{\lambda}}
& \frac{\partial ^{2} l(\alpha,\lambda)}{\partial \alpha \partial \lambda }\mid_{\hat{\alpha},\hat{\lambda}}\\
\frac{\partial ^{2} l(\alpha,\lambda)}{\partial \alpha \partial \lambda }\mid_{\hat{\alpha},\hat{\lambda}} & \frac{\partial ^{2} l(\alpha,\lambda)}{\partial \lambda^2 }\mid_{\hat{\alpha},\hat{\lambda}}  \\
\end{array} \right) \right\rbrace ^{-1}
=\left( {{\begin{array}{cc} {V^{_{11} }}& {V^{12}}  \\
		{V^{12}} & {V^{22}} \\
		\end{array}}}\right). \nonumber \\
\end{equation}

It is well known that MLEs are asymptotically normally distributed and using this property of
MLEs, we can construct the approximate confidence intervals for $ \alpha $ and $ \lambda $.
Since the $ \hat{\alpha}$ and $\hat{\lambda}$ is asymptotically normally distributed, we have the asymptotic
distribution of
$$
P_{1}=\frac{\hat{\alpha}-\alpha}{\sqrt{V^{11}}},~~~~~~~~~
P_{2}=\frac{\hat{\lambda}-\lambda}{\sqrt{V^{22}}},
$$
to be standard normal. Using the pivotal quantities $P_{1}$ and
$P_{2}$, $100(1-\gamma)\%$ asymptotic confidence intervals for $\alpha$
and $\lambda$ based on the MLEs are
\begin{equation}
(\widehat{\alpha}-z_{\frac{\gamma}{2} }\sqrt{V^{11}},
\widehat{\alpha}+z_{\frac{\gamma}{2} } \sqrt{V^{11}}),
\end{equation}
and
\begin{equation}
(\widehat{\lambda}-z_{\frac{\gamma}{2} }\sqrt{V^{22}},
\widehat{\lambda}+z_{\frac{\gamma}{2} } \sqrt{V^{22}}),
\end{equation}
respectively, where $ z_{\frac{\gamma}{2}} $ is the $ \left( \frac{\gamma}{2}\right) ^{th} $ upper percentile of standard normal distribution.

\section{Bayesian Analysis}
In this section we compute the Bayes estimates and the associated HPD credible intervals of the shape and scale parameters. We need to assume some prior distributions of the
unknown parameters for the Bayesian inference. Unfortunately, when both the parameters are unknown, there does not exist any natural conjugate priors. In this paper similarly as in  Kundu and Gupta (2008), it is
assumed that $\alpha$ and $\lambda$ have the following independent gamma priors;
$$ \pi(\alpha| a,b)\propto \alpha^{a-1} e^{-b\alpha},~~~~~~\alpha>0,  $$
$$ \pi(\lambda| c,d)\propto \lambda^{c-1} e^{-d\lambda},~~~~~~\lambda>0.$$

Here all the hyper parameters $ a, b, c, d $ are assumed to be known and positive. Based on the above priors, the joint density function of the data, $ \alpha $ and $ \lambda $ is
\begin{equation}
L(data,\alpha,\lambda)=\alpha^{r+a-1}\lambda^{r+c-1}e^{-b \alpha -\lambda \lbrace d+\sum_{i=1}^r x_i^\alpha \rbrace}\prod_{i=1}^r x_i^{\alpha+1} (1-e^{-\lambda u^{-\alpha}})^{n-r}.
\end{equation}
Based on $ L(data,\alpha,\lambda) $, we obtain the joint posterior density function of $\alpha$ and $  \lambda$ given the data as
\begin{equation}
\pi(\alpha,\lambda\mid data)=\frac{L(data,\alpha,\lambda)}
{\int_{0}^\infty \int_{0}^\infty {L(data,\alpha,\lambda)}d \alpha d\lambda}.
\end{equation}
Therefore, the posterior density function of $  \alpha $ and $  \lambda $ given the data can be written as
\begin{equation}
\pi(\alpha,\lambda\mid data)\propto g_{1}(\lambda \mid \alpha,data) g_{2}(\alpha\mid data) h(\alpha, \lambda \mid data),
\end{equation}
here $  g_{1}(\lambda| \alpha, data) $ is a gamma density function with the shape and scale parameters as $ (r+c) $ and
$ (d+\sum_{i=1}^r {x_{i}^{\alpha}}) $, respectively. Also $ g_2(\alpha\mid data) $ is a proper density function given by
\begin{equation}
g_2(\alpha\mid data)\propto \frac{1}{ (d+\sum_{i=1}^r {x_{i}^{\alpha}})^{r+c}} \alpha^{a+r-1} e^{-b\alpha} \prod_{i=1}^r {x_{i}^{\alpha+1}}. 
\end{equation}
Moreover
\begin{equation*}
h(\alpha,\lambda\mid data)=(1-e^{-\lambda u^{-\alpha}})^{n-r}.
\end{equation*}

Therefore, the Bayes estimate of any function of $  \alpha $ and $  \lambda $ , say $  g(\alpha, \lambda ) $ under the squared error loss function is
\begin{equation}\label{q55}
\hat{g}_{B}(\alpha, \lambda)=\frac{{\int_{0}^\infty \int_{0}^\infty {g(\alpha, \lambda)\ g_{1}(\lambda \mid\alpha, data) g_{2}(\alpha\mid data)}h (\alpha ,\lambda \mid data)} d\alpha d\lambda}
{{\int_{0}^\infty \int_{0}^\infty {\ g_{1}(\lambda \mid\alpha, data) g_{2}(\alpha\mid data)}h (\alpha ,\lambda \mid data)} d\alpha d\lambda}. 
\end{equation}

Unfortunately, (\ref{q55}) can not be computed analytically for general $ g(\alpha, \lambda )$. We apply two different approximation methods to evaluate the Bayes estimators of $ \alpha $ and $  \lambda $. The first approximation
technique due to Lindley (1980) and the second is an importance sampling procedure as suggested by Chen and Shao (1999). The details are explained below.

\subsection{Lindley's Approximation}
It is known that the (\ref{q55}) can not be computed explicitly. Because of that Lindley(1980) proposed an approximation to compute the ratio of two integrals such as (\ref{q55}). This has been used by several authors to obtain the approximate Bayes estimators. This approximation technique uses Taylor's series expansion of the integral expression around maximum likelihood estimator.

Based on Lindley's approximation, the approximate Bayes estimates of $ \alpha $ and $ \lambda $ under the squared error loss functions are respectively

\begin{align}
\hat {\alpha}_{L}&= \hat{\alpha}+\frac{1}{2}\Bigg [  l _{30}
\tau_{11}^{2}+l_{03}\tau_{21}\tau_{22}+3
l_{21}\tau_{11}\tau_{12}+l_{12}(\tau_{22}\tau_{11}+2\tau_{21}^{2})\Bigg]\nonumber\\
&+\left( \frac{a-1}{\hat{\alpha}}
-b\right) \tau_{11} +\left( \frac{c-1}{\hat{\lambda}} -d\right) \tau_{12}\\
\hat {\lambda}_{L}&=\hat{\lambda}+\frac{1}{2}\Bigg [  l _{30}
\tau_{12}\tau_{11}+l_{03}\tau_{22}^{2}+l_{21}
(\tau_{11}\tau_{22}+2\tau_{12}^{2})+3
l_{12}\tau_{22}\tau_{21}\Bigg]\nonumber\\
&+\left( \frac{a-1}{\hat{\alpha}} -b\right) \tau_{21} +\left(
\frac{c-1}{\hat{\lambda}} -d\right) \tau_{22},
\end{align}
where $ \hat{\alpha} $  and  $ \hat{\lambda} $ are MLEs of $ \alpha $
and $ \lambda $ respectively and a, b, c, d are the known hyper
parameters. The explicit expressions of $
\tau_{11},\tau_{12},\tau_{22},l_{30},\tau_{21}, l_{03}, l_{12},
l_{21}$ are provided in the Appendix A. \

Although using Lindley's approximation we can obtain the Bayes estimates, but it is not possible to construct the HPD credible intervals using this method. Therefore, in the next subsection we propose the importance sampling procedure to draw samples from the posterior density function and in turn compute the Bayes estimates, and also construct HPD credible intervals.

\subsection{Importance sampling}
 We use importance sampling to generate a sample from the posterior density function $\pi(\alpha, \lambda \mid data)$ and then to compute the Bayes estimates and HPD credible intervals. The following theorem can be useful for further development.

\begin{theorem}
The conditional density of $  \alpha$, given data, say $ g_2(\alpha\mid data) $ is log-concave.
\end{theorem}

\begin{theproof}
 See Appendix B.
\end{theproof}

Because of log-concavity of $ g_2(\alpha\mid data)$, the idea of Devroye (1984) can be used to generate a sample from $ g_2(\alpha\mid data)$. Moreover, since $ g_1(\lambda \mid \alpha, data)$ follows gamma, it is quite simple to generate from  $ g_1(\lambda \mid \alpha, data)$. Now we would
like to provide the importance sampling procedure to compute the Bayes estimates and also to construct the credible interval of  $ g(\alpha, \beta)=\theta $ (say). Similarly as in Raqab and Madi(2005) a simulation based consistent estimate of $ E( g(\alpha, \beta))=E(\theta) $ can be obtained using Algorithm as given below.
\begin{flushleft}
	\textbf{Algorithm. }
\end{flushleft}
\begin{itemize}
	\item [Step 1:] Generate $ \alpha $ from $ g_2(.\mid data)$ using the method developed by Devroye (1984).
	\item [Step 2:] Generate $ \lambda $ from $ g_1(. \mid \alpha, data)$.
	\item [Step 3:] Repeat Step 1 and Step 2 and obtain $ (\alpha_{1}, \lambda_{1}), ..., (\alpha_{M},\lambda_{M} ) $.
	\item [Step 4:]: An approximate Bayes estimate of $  \theta $ under a squared error loss function can be obtained as
	\begin{equation}
	\hat{g}_{B}(\alpha, \lambda)=\hat{\theta}=\frac{\frac{1}{M}\sum_{i=1}^M{\theta _i h(\alpha_i , \lambda_i \mid data)}}{\frac{1}{M}\sum_{i=1}^M{ h(\alpha_i , \lambda_i \mid data)}}.
	\end{equation}
	\item [Step 5:] Obtain the posterior variance of  $ g(\alpha, \beta)=\theta $ as
	\begin{equation}
	\hat{V}(g(\alpha, \lambda\mid data))=\frac{\frac{1}{M}\sum_{i=1}^M{(\theta _i-\hat{\theta})^{2} h(\alpha_i , \lambda_i \mid data)}}{\frac{1}{M}\sum_{i=1}^M{ h(\alpha_i , \lambda_i \mid data)}}.
	\end{equation}
\end{itemize}

We now obtain the credible interval of $ \theta $ using the idea of Chen and Shao (1999). Let us denote $ \pi(\theta \mid data)$ and
$ \Pi(\theta \mid data)$ as the posterior density and posterior distribution functions of $ \theta $, respectively. Also let $\theta^{(\beta)}$ be the $ \beta$-th quantile of $\theta$, i.e.,
\begin{equation*}
\theta^{(\beta)}=inf\lbrace\theta; \Pi(\theta\mid data)\geq\beta\rbrace, ~~~~~0<\beta <1.
\end{equation*}
Observe that for a given $ \theta^{*} $,
$ \Pi(\theta^{*} \mid data)=E\lbrace I_{\theta\leq\theta^{*}}\mid data\rbrace$, where $ I_{\theta\leq\theta^{*}} $ is the indicator function.
Therefore, a simulation consistent estimator of $ \Pi(\theta^{*} \mid data)$ can be obtained as
\begin{equation}
\hat{\Pi}(\theta^{*} \mid data)=\frac{\frac{1}{M}\sum_{i=1}^M{I_{\theta_{i} \leq \theta^{*}} h(\alpha_i , \lambda_i \mid data)}}{\frac{1}{M}\sum_{i=1}^M{ h(\alpha_i , \lambda_i \mid data)}}.
\end{equation}

For $ i=1,...,M $, let $ \lbrace \theta_{(i)}\rbrace $ be the ordered values of $ \theta_{i} $, and $$ w_{i}=\frac{h(\alpha_{(i)} , \lambda_{(i)} \mid data)}{\sum_{i=1}^M{ h(\alpha_i , \lambda_i \mid data)}},$$
be the associated weight, then we have
\begin{equation}
\hat{\Pi}(\theta^{*} \mid data)=
\left\{\begin{array}{l}
0~~~~~~~~~~~~~~~~ \theta^{*}<\theta_{(1)} \\
\sum_{j=1}^{i}{w_{j}}~~~~~ \theta_{(i)} \leq\theta^{*}<\theta_{(i+1)} \\
1~~~~~~~~~~~~~~~~\theta^{*}\geq\theta_{(M)}.
\end{array}\right.
\end{equation}
Therefore, $ \theta^{(\beta)} $ can be approximated by
\begin{equation}
\hat{\theta}^{(\beta)}=
\left\{\begin{array}{l}
\theta_{(1)}~~~~~~~~~~~~ \beta =0 \\
\theta_{(i)}~~~~~~~~~~~\sum_{j=1}^{i-1}{w_{j}}<\beta\leq
\sum_{j=1}^{i}{w_{j}}~~~~~ .
\end{array}\right.
\end{equation}

To obtain a $100(1-\beta)\%$ HPD credible interval for $\theta $,
consider intervals of the form
\begin{equation}
R_{j}= \left( \hat{\theta}^{\left(\frac{j}{M}\right)}  , \hat{\theta}^{\left(\frac{j+[(1-\beta)M]}{M}\right)}\right)
\end{equation}
for $ j=1,2,\ldots,M-[(1-\beta)M]$, where $[a]$ denotes the largest integer less than or equal to  $[a]$. Finally, among all $R_{j}$ choose that interval which has the smallest length.

\section{Simulation Results}
In this section we compare the performance of the different methods through a simulation study. We estimate the unknown parameters using the MLE, Bayes estimators obtained by Lindley's approximations, and also by the Bayes estimators obtained by using MCMC technique. The simulation study is carried out for different sample size and with different choices of  $R$, $T$ values. For a particular set of hybrid censored data, the MLEs and Bayes estimators are obtained as described before. Both non-informative and informative priors are used for the shape and scale parameters. In case of non-informative prior we take $ a = b = c = d = 0 $. We call it as Prior 1. Note that as the hyper-parameters go to zero, the prior density becomes inversely proportional to its argument and also becomes improper. This density is commonly used as an improper prior for parameters in the range of zero to infinity. It should also be mentioned that when $ a = b=0 $ ,  $ \pi(\alpha| a,b) $  is not log-concave, but the posterior density function $g_2(\alpha\mid data)$ is still log-concave. For the informative
prior, we chose $a = 2, b = 1, c = d = 1 $. We call it as Prior 2. For computing different point estimators we generated 1000 samples from the IW distribution with $ \alpha=2 $ and $\lambda=1 $. The averages and mean squared
errors (MSE) of estimators of  $\alpha $ and $\lambda $ are presented in Tables 1 and 2, respectively. \\
We also compute the 95\% asymptotic confidence intervals based on MLEs. For comparison purposes, we compute the $95\%$ HPD credible intervals from the Gibbs samples. We report the average confidence/credible lengths in Table 3.
In Table 3, the first and second row represent the result for $\alpha $ and $\lambda $, respectively. \\
Some of the points are quite clear from Tables 1 and 2. In Tables 1 and 2, it is observed that the approximate Bayes estimators of unknown parameters based on Lindley's approximation match quite well with the Bayes estimators using MCMC method.

\begin{table}[t]
	\caption{ The average estimates (A.E) and the associated MSE for $ \alpha $.}
	\centering
	\footnotesize
	\begin{tabular}{|c|c|c|c|c|c|}
		
		\hline
		~~~$ ( n ,T ) $ ~~~  & ~~$ R $ ~~&~~~~ & ~~~MLE~~~& ~~~ Bayes(Lindley) ~~~ &~~~Bayes(MCMC) ~~~   \\
		\cline{5-6}
		&       &      &    &  prior 1 ~~~ prior 2 & prior 1 ~~~ prior 2     \\
		\hline
		& 20 &  A.E  &   2.1113  & 2.1789~~~2.1068 & 2.1249  ~~~ 2.0919  \\
		&     & MSE    &  0.1848   &0.2078~~~0.1646 &    0.1325  ~~~  0.0953 \\
		\cline{2-6}
		(30, 1.5) &  25   &  A.E  &  2.1091 &2.0665~~~ 2.0948 &  2.1119  ~~~  2.0992  \\
		&    & MSE    &  0.1503  & 0.1382~~~0.1035 &  0.1463  ~~~   0.1210  \\
		\cline{2-6}
		&  30  & A.E  &  2.1057  &2.1356 ~~~2.0855 &    2.1062  ~~~ 2.0826  \\
		&           & MSE    & 0.1519  &0.1638 ~~~ 0.1231 & 0.1211   ~~~  0.1163  \\
		\hline
		& 20  & A.E  & 2.1394 &2.1244 ~~~ 2.0993 &  2.1253 ~~~ 2.1256 \\
		&           & MSE    & 0.1696  &0.2187 ~~~ 0.1453 & 0.1218 ~~~ 0.0671 \\
		\cline{2-6}
		(30, 2.5) & 25  & A.E &  2.0999 &2.1224~~~ 2.0851 &  2.0687~~~ 2.0639 \\
		&         & MSE    & 0.1317 & 0.1708 ~~~ 0.1287 & 0.1054  ~~~  0.0261  \\
		\cline{2-6}
		& 30  & A.E  &  2.0996 & 2.0466~~~2.1142& 1.9790 ~~~1.9985 \\
		&              & MSE    &  0.1306  &0.3042~~~ 0.1625 &   0.1142  ~~~ 0.0184   \\
		\hline
		& 35 &  A.E  &  2.0586  &2.0540 ~~~ 2.0209&   2.0495 ~~~ 2.0262  \\
		&     & MSE    & 0.0820   &0.0740~~~0.0727&    0.0667  ~~~   0.0318  \\
		\cline{2-6}
		(50,1.5) &  40  & A.E  & 2.0633 &2.0738 ~~~ 2.0604 &  2.0424 ~~~ 1.9912 \\
		&         & MSE    &  0.0791  &0.1017 ~~~ 0.0752&  0.0520  ~~~ 0.0275  \\
		\cline{2-6}
		& 50  & A.E  &   2.0623  &2.0822~~~ 2.0609 &   2.0136  ~~~ 2.0124  \\
		&           & MSE    & 0.0764    &0.0847~~~ 0.0809&    0.0709   ~~~  0.0152  \\
		\hline
		& 35  & A.E  &  2.0793 & 2.0795 ~~~ 2.0430 & 2.0196 ~~~2.0149 \\
		&           & MSE    & 0.0747 &0.0741~~~ 0.0649 & 0.0481 ~~~ 0.0032 \\
		\cline{2-6}
		(50,2.5)  & 40  & A.E  &  2.0682 &2.0799~~~ 2.0847 & 2.0085~~~2.0068\\
		&         & MSE    &     0.0737  &0.0722 ~~~0.0839&  0.0296 ~~~ 0.0017  \\
		\cline{2-6}
		& 50  & A.E  &    2.0287  & 2.0525~~~2.0376&  2.0013  ~~~ 2.0008   \\
		&      &MSE   &  0.0691    &0.0891~~~ 0.0547& 0.0154  ~~~   0.0006  \\
		\hline
	\end{tabular}
\end{table}

\begin{table}[ht]
	\caption{ The average estimates (A.E) and the associated MSE for $ \lambda $.}
	\centering
	\footnotesize
	\begin{tabular}{|c|c|c|c|c|c|}
		
		\hline
		~~~$ ( n ,T ) $ ~~~  & ~~$ R $ ~~&~~~~ & ~~~MLE~~~& ~~~ Bayes(Lindley) ~~~ &~~~Bayes(MCMC) ~~~   \\
		\cline{5-6}
		&       &      &    &  prior 1 ~~~ prior 2 & prior 1 ~~~ prior 2     \\
		\hline
		
		& 20 &  A.E  & 1.0464  & 0.9793  ~~~ 1.0119  & 1.0570  ~~~ 1.0537 \\
		(30,1.5) &     & MSE    & 0.0514  & 0.0456~~~0.0472&  0.0528  ~~~ 0.0425  \\
		\cline{2-6}
		&  25   &  A.E  &  1.0298  &1.0164~~~1.0170 & 0.9971  ~~~  0.9993 \\
		&    & MSE    &   0.0487  &0.0406~~~ 0.0335&  0.0452  ~~~   0.0418  \\
		\cline{2-6}
		&  30  & A.E &  1.0264  & 1.0122~~~1.0216&  1.0371   ~~~ 1.0283  \\
		&           & MSE    & 0.0461   &0.0526~~~ 0.0418 &  0.0487  ~~~   0.0395 \\
		\hline
		& 20  & A.E  & 1.0107& 1.0326~~~1.0071& 1.0203 ~~~  1.0237  \\
		(30,2.5) &       & MSE    &  0.0467 &0.0401~~~ 0.0445& 0.0392~~~  0.0359 \\
		\cline{2-6}
		& 25  & A.E  & 1.0275 &1.0055 ~~~1.0449 & 1.0159 ~~~ 0.9787 \\
		&         & MSE    &  0.0451 &0.0591 ~~~ 0.0774 &  0.0309 ~~~ 0.0287  \\
		\cline{2-6}
		& 30  & A.E &  1.0282 &1.0673~~~1.0342 &  0.9958 ~~~ 0.9864 \\
		&              & MSE    &  0.0437 &0.0504~~~0.0418 &   0.0377 ~~~  0.0261  \\
		\hline
		& 35 &  A.E  &  1.012 &1.0002~~~1.0167& 1.0423  ~~~  1.0337  \\
		(50,1.5) &     & MSE    & 0.0284   &0.0258 ~~~ 0.0229&   0.0276 ~~~  0.0246  \\
		\cline{2-6}
		&  40   &  A.E   &  1.0153  &0.9944~~~1.0144 &  1.0334  ~~~ 1.0221   \\
		&    & MSE    &    0.0271  &0.0243 ~~~ 0.0226 &   0.0262  ~~~  0.0241  \\
		\cline{2-6}
		&  50  & A.E   &   1.0267  &1.0131~~~1.0175&   1.017  ~~~ 1.0263  \\
		&           & MSE    & 0.0266   &0.0230 ~~~ 0.0278 &   0.0243  ~~~   0.0240 \\
		\hline
		& 35  & A.E   & 1.0159 &1.0032 ~~~ 1.0013& 1.0446 ~~~ 1.0373\\
		(50,2.5) &     & MSE & 0.0260   & 0.0287~~~0.0259&  0.0216 ~~~ 0.0207 \\
		\cline{2-6}
		& 40  & A.E   & 1.0221 &1.0037 ~~~ 0.9903 & 1.0245 ~~~ 1.0327\\
		&         & MSE    &   0.0280  &0.0255 ~~~ 0.0482 &  0.0219  ~~~  0.0135  \\
		\cline{2-6}
		& 50  & A.E    &  1.0228 &1.0322 ~~~ 1.0253& 1.0087  ~~~ 1.0074  \\
		&       & MSE   &  0.0262  &0.0416 ~~~ 0.0261&   0.0204  ~~~ 0.0018  \\
		\hline
	\end{tabular}
\end{table}

 In most of the cases, the Bayes estimates  obtained by using Lindley's approximation of $ \lambda $ based on prior 1 perform
better than the MLEs of $\lambda  $, but while for $ \alpha $ it is the other way. But in case of prior 2, the Bayes estimates using Lindley's approximation of ($\alpha, \lambda $) perform marginally better than the MLEs for all cases considered.  It is also observed that in most of cases the performance in terms of average bias and the MSE of Bayes estimates obtained by using MCMC procedure under Prior 1 are close to that of the corresponding behaviour of the MLEs or the Bayes estimates obtained by Lindley's approximations. But while using informative prior (Prior 2), the performance of the Bayes estimates by using MCMC are much better than the other estimates. Therefore, if the prior information are available, then we should use the Bayes estimates, otherwise MLEs may be used to avoid the computational cost.\

For all the methods, and for both the estimators, it is observed that
for fixed $ n $ as $ R $ or $ T $ increases in most of cases the average biases,
and the MSE decrease, it verifies the consistency properties of the estimates.\

Now, considering the confidence intervals and credible lengths, it is observed that the asymptotic results of the MLE
work quite well. It can be seen that the average confidence lengths is quite
close to the average credible intervals, mainly for large $ n $ and $ R $. But, in most of the cases,
the average lengths of the credible intervals are slightly shorter than the confidence intervals. From Table 3 it is observed that the results obtained using informative priors are not significantly different than the corresponding results obtained using non-proper priors.  Finally, note that Bayes estimates are most computationally expensive, followed by MLE.

\begin{table}[ht]
	\caption{ The average confidence/credible lengths for the MLE and Bayes estimates of $ \alpha $ and $ \lambda $.}
	\centering
	\footnotesize
	\begin{tabular}{|c|c|c|c|}
		
		\hline
		~~~~$ ( n ,T ) $~~~~  & ~~~~$ R $ ~~~~ & ~~~~~MLE~~~~~ & Bayes(MCMC) \\
		\cline{4-4}
		&        &      & prior 1 ~~~ prior 2      \\
		\hline
		&  20  & 1.4718  &   1.3986   ~~~  1.3864  \\
		&        & 0.8106  & 0.8230   ~~~  0.79180  \\
		\cline{2-4}
		(30,1.5) & 25  &   1.4338  &  1.3891   ~~~  1.3828  \\
		&    &  0.7978   &   0.7663   ~~~   0.7565  \\
		\cline{2-4}
		&  30  &   1.4422  &  1.3911    ~~~ 1.3872 \\
		&         & 0.7996   &   0.7508    ~~~   0.7500  \\
		\hline
		
		& 20    & 1.4384 & 1.4024  ~~~ 1.3835 \\
		&        & 0.7891  &  0.8013  ~~~ 0.7812 \\
		\cline{2-4}
		(30,2.5) & 25    &  1.2730 & 1.1916 ~~~ 1.2014 \\
		&         &  0.7873  &  0.7818  ~~~  0.7416  \\
		\cline{2-4}
		& 30  &  1.2533 & 1.1885 ~~~ 1.1762\\
		&           &    0.7788  &  0.7417  ~~~  0.7469  \\
		\hline
		
		& 35 &    1.1985  & 1.1842   ~~~  1.1873  \\
		&      &   0.6144 &   0.6327  ~~~   0.5990  \\
		\cline{2-4}
		(50,1.5) &  40   &   1.0917  & 1.0916  ~~~ 1.0913\\
		&    &   0.6147  &  0.5986  ~~~   0.5295 \\
		\cline{2-4}
		&  50   & 1.0889  &   1.0862   ~~~1.0885  \\
		&         &  0.6164 &   0.6069    ~~~  0.6014  \\
		\hline
		
		& 35   &  1.0450 & 1.0381 ~~~ 1.0399 \\
		&     &  0.6022 & 0.5499 ~~~ 0.5427 \\
		\cline{2-4}
		(50,2.5)& 40   & 0.9934 & 1.0057 ~~~ 0.9831\\
		&         &  0.5968  &  0.5259   ~~~ 0.5082  \\
		\cline{2-4}
		& 50    &  0.9483 & 0.9129  ~~~ 0.9210 \\
		&         &   0.5958  &   0.5024  ~~~  0.5169   \\
		\hline
	\end{tabular}
\end{table}

\section{Illustration}
In this section, we consider the two following examples to illustrate the use of the estimation methods proposed in the previous sections.\\

{\bf Example 1.}
Consider the following data giving the maximum flood levels (in
millions of cubic feet per second) of the Susquehenna River at
Harrisburg, Pennsylvenia over 20 four-year periods (1890 $-$ 1969).
These data are taken from Dumonceaux and Antle (1973).

\begin{center}
	\begin{verbatim}
   0.654  0.613  0.315  0.449  0.297  0.402  0.379  0.423  0.379  0.324  
	   0.269  0.740  0.418  0.412  0.494  0.416  0.338  0.392  0.484  0.265 
	\end{verbatim}
	
\end{center}

\begin{figure}[t]
	\begin{center}
		\includegraphics[width=10cm,height=8cm]{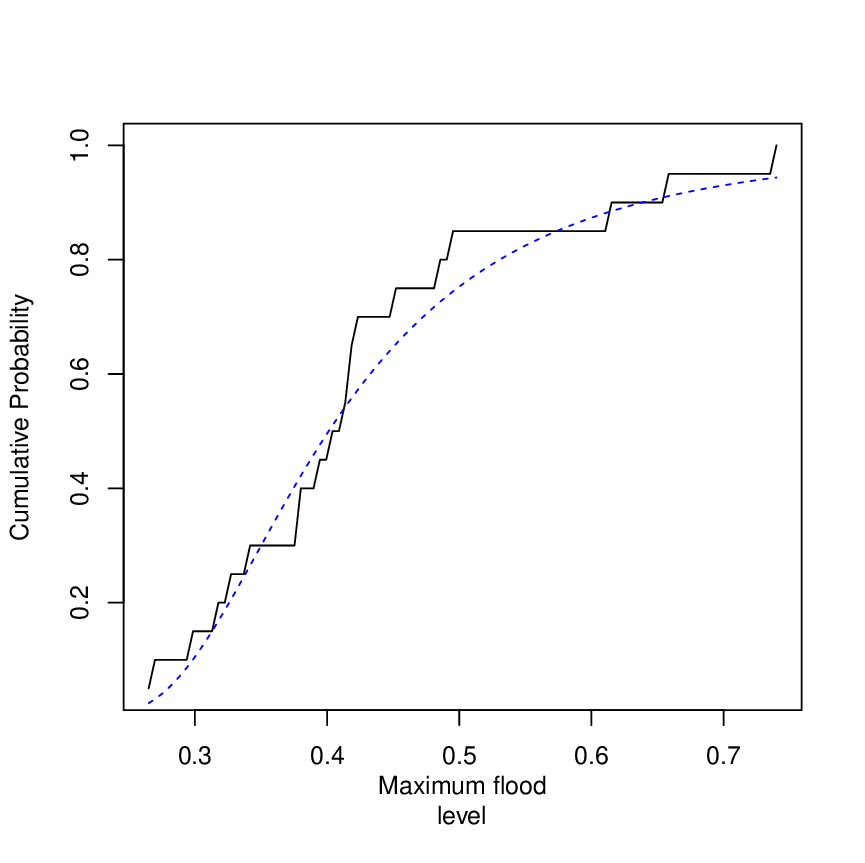}
		\caption{\footnotesize{The empirical and fitted distributions.}}\label{fig1}
	\end{center}
\end{figure}
Before progressing, first we want to check whether the IW distribution fits the data or not. For this purpose, we have used the complete data. The MLEs and bayes estimates of  ($\alpha $,$ \theta $) based on the complete sample are (4.3143, 2.7905) and (4.1861, 2.7657), respectively.
The Kolmogorov-Smirnov distance between the empirical distribution function and the fitted distribution functions when the parameters are obtained by MLEs, and the associated $p$-value are 0.1060 and 0.8557, respectively. Since the $p$-value is quite high, we cannot reject the null hypothesis that the data are coming from the IW distribution.
We have just plotted the empirical cumulative distribution function and the fitted cumulative distribution function in Figure 1. It shows that IWD fit the data very well. Now, we have created two artificially the Type-I hybrid censored data sets from the above uncensored data set, using the following censoring schemes:\\

\textbf{Scheme 1:} $ R = 18, T =0.5. $ \\
In this scheme, it is observed that the $ R $-th failure does not take place before time point $ T $. For this scheme, the hybrid censored sample is:
\begin{center}
	\begin{verbatim}
    0.265  0.269  0.297  0.315  0.324  0.338  0.379  0.379  0.392  0.402 
	    0.412  0.416  0.418  0.423  0.449  0.484  0.494
	\end{verbatim}
\end{center}

From the above sample data, the MLEs of $  \alpha$ and $  \theta $ are   4.2726  and  2.6565, respectively. Since we do not have any prior information available, we use non-informative priors, i.e., $ a = b = c = d = 0 $ on both $ \alpha $ and  $ \theta $ to compute bayes estimatores. Now using Algorithm of section 4.2, we generate 1000 MCMC samples and based on them we compute the Bayes estimates of $ \alpha $ and $  \theta $ as 4.5665 and 2.8148, respectively. The 95\% asymptotic confidence intervals of $  \alpha$ and $ \theta $ based on the empirical Fisher information matrix are (2.7207, 5.8244) and (2.3623,  2.9507) respectively. Moreover, the 95\% HPD credible intervals of $ \alpha $ and $ \theta $ are (2.4603, 5.2454), and (2.2977, 2.8928), respectively.\

\textbf{Scheme 2}: $ R =14, T = 0.45. $
In this scheme, it is observed that the $ R $-th failure took place before $ T $. In this case, the hybrid sample
is:
\begin{center}
	\begin{verbatim}
   0.265  0.269  0.297  0.315  0.324  0.338  0.379  0.379  0.392  0.402
	   0.412  0.416  0.418  0.423
	\end{verbatim}
\end{center}

Based on the sample, the MLEs and bayes estimates of $ \alpha$ and $ \theta $ are (3.6933,  2.5446)  and   (3.8607,  2.7158) respectively. The 95\% asymptotic confidence intervals of $  \alpha $ and $  \theta $ are (2.2233, 5.1635) and (2.2088, 2.8804) respectively. We also compute the 95\% HPD credible intervals of $ \alpha $ and $ \theta $ and they are (2.3481, 5.2153), and (2.2718, 3.0145), respectively.\\

{\bf Example 2.}
In this example we consider the data given by Bjerkedal (1960), and it represents the survival times (in days) of guinea pigs injected with different doses of tubercle bacilli. The regimen number is the common logarithm of the number of bacillary units in 0.5 ml. of challenge solution; i.e., regimen 6.6 corresponds to $4.0\times10^6$ bacillary units per 0.5 ml. (log $(4.0 \times 10^{6})= 6.6).$ Corresponding to regimen 6.6, there were 72 observations listed below:
\begin{center}
	\begin{verbatim}
	12  15  22  24  24  32  32  33  34  38  38  43  44  48  52  53  54  54  
	55  56  57  58  58  59  60  60  60  60  61  62  63  65  65  67  68  70  
	70  72  73  75  76  76  81  83  84  85  87  91  95  96  98  99  109 110  
	121 127 129 131 143 146 146 175 175 211 233 258 258 263 297 341 341 376.
	\end{verbatim}
\end{center}

Kundu and Howlader (2010) indicate that  IW model with
$\alpha=1.4142 $ and $\theta=0.0169$  is suitable for this data. For illustrative purposes, we have generated two different Type-I hybrid censored samples using the following two sampling schemes:\\
\textbf{Scheme 1}: $ R = 50, T =90 $ and \textbf{Scheme 2}: $ R =60, T = 150. $\
For scheme 1, the hybrid censored sample is:\
\begin{center}
	\begin{verbatim}
	  12  15  22  24  24  32  32  33  34  38  38  43  44  48  52  53  54
  54  55  56  57  58  58  59  60  60  60  60  61  62  63  65  65  67
	  68  70  70  72  73  75  76  76  81  83  84  85  87.
	\end{verbatim}
\end{center}

For this data set, the MLEs of $  \alpha$ and $  \theta $ are  1.3272  and 0.0178 respectively. We compute the Bayes estimates of $ \alpha $ and $  \theta $ with respect to the assumed non-informative priors as 1.4736 and 0.0142, respectively. The 95\% asymptotic confidence intervals of $  \alpha$ and $ \theta $ are (1.0569, 1.5779) and (0.0140,0.0212) respectively. Similarly, the 95\% HPD credible intervals of $ \alpha $ and $ \theta $ are (1.2604, 1.6868), and (0.0106, 0.0177), respectively.

Now for Scheme 2, it is observed that the $ R $-th failure took place before $ T $. In this case, the hybrid sample
is:\
\begin{center}
	\begin{verbatim}
	  12  15  22  24  24  32  32  33  34  38  38  43  44  48  52  53  54 
	  54  55  56  57  58  58  59  60  60  60  60  61  62  63  65  65  67
	  68  70  70  72  73  75  76  76  81  83  84  85  87  91  95  96  98  
	  99  109 110 121 127 129 131 143 146.
	\end{verbatim}
\end{center}

Based on the sample, the MLEs and bayes estimatores of $ \alpha $ and $  \theta $ are (1.3688,  0.0182)  and   (1.4599,  0.0158) respectively. The 95\% asymptotic confidence intervals of $  \alpha $ and $  \theta $ are (1.1846, 1.6447) and (0.0152, 0.0216) respectively. We also compute the 95\% HPD credible intervals of $ \alpha $ and $ \theta $ as (1.3426, 1.5763) and (0.0135, 0.0181), respectively.

\section{Conclusion}

In this paper we considered the classical and Bayesian inference of the inverse Weibull distribution based on Type-I hybrid censored data. The maximum likelihood estimators of the parameters can be obtained by using an iterative procedure. Hence the Bayesian inference seems to be the natural choice for the analysis of certain survival data. The prior belief of the model was represented by the independent gamma priors on the shape and scale parameters. The squared error loss function was used as it is appropriate when large errors of the estimation are considered to be more serious compared to other loss functions.
It was observed that the Bayes estimators and the HPD credible intervals can not be obtained in explicit form. We proposed two approximations which can be implemented very easily. We compared the performance of the Bayes estimators with the MLEs by Monte Carlo simulations, and it was observed that the performances are quite satisfactory.



\section*{Appendix A}
For the two-parameter case, using the notation $(\lambda_{1}, \lambda_{2})=(\alpha, \lambda)  $, the Lindley's approximation can be written as,

$$
\hat {g}=g(\hat{\lambda}_{1}, \hat{\lambda}_{2})+\frac{1}{2}\Bigg [A+  l _{30}B_{12}+l_{03}B_{21}+l_{21}C_{12}+ l_{12}C_{21} \Bigg]+ p_{1}A_{12}+p_{2}A_{21},
$$

where,
$$A= \sum_{i=1}^2{\sum_{j=1}^2{w_{ij}}\tau_{ij}},~~~~ l_{ij}=\frac{\partial^{i+j}L(\lambda _{1},\lambda_{2})}{\partial  \lambda_{1}^{i}\partial  \lambda_{2}^{j}}, i,j=0,1,2,3, i+j=3 ,~~~ p_{i}=\frac{\partial p}{\partial \lambda_{i}},
$$

$$
w_{i}=\frac{\partial g}{\partial \lambda_{i}},~~ w_{ij}=\frac{\partial^{2} g}{\partial \lambda_{i} \partial \lambda_{j}},~~~ p=\ln \pi(\lambda_{1}, \lambda_{2}),~~~ A_{ij}=w_{i}\tau_{ii}+w_{j}\tau_{ji}
$$
$$
B_{ij}=(w_{i}\tau_{ii}+w_{j}\tau_{ij})\tau_{ii}, C_{ij}=3w_{i}\tau_{ii}\tau_{ij}+w_{j}(\tau_{ii}\tau_{jj}+2 \tau_{ij}^{2}),
$$
here L(.,.) is the log-likelihood function of the observed data, $ \pi(\lambda_{1}, \lambda_{2})  $  is the joint prior density function of $(\lambda_{1}, \lambda_{2})  $ and $ \tau_{ij} $ is the (i,j)-th element of the inverse of the observed Fisher information matrix. Moreover, $ \hat{\lambda}_{1} $ and  $ \hat{\lambda}_{2} $ are the MLEs of $(\lambda_{1}$, $\lambda_{2}) $, respectively and all the quantities are evaluated at ( $ \hat{\lambda}_{1} $,  $ \hat{\lambda}_{2} $).\\
\hspace{0.5cm}Now we have
$$
L(\alpha,\lambda\mid x)=\alpha^{r}\lambda^{r}e^{-\lambda \sum_{i=1}^r x_i^\alpha}\prod_{i=1}^r x_i^{\alpha+1} (1-e^{-\lambda u^{-\alpha}})^{n-r}.
$$
Therefore, we obtain
\begin{align*}
l_{03}&= \frac{2r}{\hat {\lambda}^3}+\frac{(n-r) u^{-3\hat {\alpha}} e^{-\hat {\lambda} u^{-\hat {\alpha}}}}{1-e^{-\hat {\lambda} u^{-\hat {\alpha}}}}+\frac{3(n-r) u^{-3\hat {\alpha}} e^{-2\hat {\lambda} u^{-\hat {\alpha}}}}{{(1-e^{-\hat {\lambda} u^{-\hat {\alpha}}})}^2}+\frac{2(n-r) u^{-3\hat {\alpha}} e^{-3\hat {\lambda} u^{-\hat {\alpha}}}}{{(1-e^{-\hat {\lambda} u^{-\hat {\alpha}}})}^3},\\
l_{30}&=\frac{2r}{\hat {\alpha}^3}-\frac{(n-r)\hat {\lambda} u^{-\hat {\alpha}}(\ln u)^3 e^{-\hat {\lambda} u^{-\hat {\alpha}}}(1-3\hat {\lambda} u^{-\alpha}+\hat {\lambda}^{2}u^{-2\alpha})}{1-e^{-\hat {\lambda} u^{-\alpha}}}\nonumber\\
&+\frac{3(n-r)\hat {\lambda}^{2} u^{-2\hat {\alpha}} \ln u^{3} e^{-2\hat {\lambda} u^{- \hat {\alpha}}}(1-\hat {\lambda} u^{-\hat {\alpha}})}{{(1-e^{-\hat {\lambda} u^{-\hat {\alpha}}})}^2}-\hat {\lambda}\sum_{i=1}^r {x_{i}^{\hat {\alpha}}( \ln x_{i})^3}\nonumber\\
&-\frac{2(n-r) \hat {\lambda}^{3} u^{-3\hat {\alpha}} \ln u^{3} e^{-3\hat {\lambda} u^{- \hat {\alpha}}}(1-\hat {\lambda} u^{-\hat {\alpha}})}{{(1-e^{-\hat {\lambda} u^{-\hat {\alpha}}})}^3},\\
l_{12}&= \frac{(n-r) u^{-2 \hat {\alpha}} \ln (u) e^{-\hat {\lambda} u^{-\hat {\alpha}}}(2-\hat {\lambda} u^{-\hat {\alpha}})}{1-e^{-\hat {\lambda}u^{-\hat {\alpha}}}}-\frac{2(n-r)\lambda u^{-3\hat {\alpha}} \ln(u)e^{-3\hat {\lambda} u^{- \hat {\alpha}}}}{{(1-e^{-\hat {\lambda} u^{-\hat {\alpha}}})}^3}\nonumber\\
&+\frac{(n-r) u^{-2\hat {\alpha}} \ln(u) e^{-2\hat {\lambda} u^{-\hat {\alpha}}}(2-3\hat {\lambda} u^{-\hat {\alpha}})}{{(1-e^{-\hat {\lambda} u^{-\hat {\alpha}}})}^2},
\end{align*}
\begin{align*}
l_{21}&=-\sum_{i=1}^r {x_{i}^{\hat {\alpha}}( \ln x_{i})^2}+\frac{(n-r) u^{-\hat {\alpha}}(\ln u)^2 e^{-\hat {\lambda} u^{-\hat {\alpha}}}(1-3\hat {\lambda} u^{-\hat {\alpha}}+\hat {\lambda}^{2}u^{-2\hat {\alpha}})}{1-e^{-\hat {\lambda}u^{-\hat {\alpha}}}}\nonumber\\
&-\frac{3(n-r) \hat {\lambda} u^{-2\hat {\alpha}} \ln(u)^{2} e^{-2\hat {\lambda} u^{-\hat {\alpha}}}(1-\hat {\lambda} u^{-\hat {\alpha}})}{{(1-e^{-\hat {\lambda} u^{-\hat {\alpha}}})}^2}\nonumber\\
&+\frac{2(n-r)\hat {\lambda}^{2} u^{-3\hat {\alpha}} \ln u^{2} e^{-3\hat {\lambda} u^{- \alpha}}}{{(1-e^{-\hat {\lambda} u^{-\hat {\alpha}}})}^3}.\
\end{align*}
\\
The elements of the Fisher information matrix are obtained in section 4. By using (\ref{qq1}), $\tau_{ij}=V^{ij}$, where $ i,j=1,2.$  \\

Now when $ g(\alpha,\lambda)=\alpha$, then

$$ w_1 = 1, w_2 = 0,~~ w_ij = 0 ,~~\textrm{for}~~i, j = 1,2. $$
Therefore,
$$
A= 0 , B_{12}=\tau_{11}^{2}, B_{21}=\tau_{21}\tau_{22},  C_{12} = 3\tau_{11}\tau_{12},
$$
$$
C_{21} = (\tau_{22} \tau_{11}+2\tau_{21}^{2}),  A_{12} = \tau_{11}, A_{21} = \tau_{12}.
$$

Now the first part of Lindley’s approximation follows by using \

$$p_1 =\frac{d-1}{\alpha} -c , p_1 =\frac{b-1}{\lambda} -a.$$

For the second part, note that $g(\alpha,\lambda)=\lambda  $, then \

$$ w_1 =0, w_2 = 1,  w_ij = 0 ,~~\textrm{for}~~i, j = 1,2.$$ \\
Therefore,
$$
A= 0 , B_{12}=\tau_{12}\tau_{11}, B_{21}=\tau_{22}^{2}, C_{12} = (\tau_{11}\tau_{22}+2\tau_{12}^{2})
$$
$$
C_{21} = 3\tau_{22}\tau_{21}, A_{12} = \tau_{21}, A_{21} = \tau_{22}.
$$
therefore the second part follows immediately.

\section*{Appendix B}
The conditional density of $\alpha$ given the data is
$$
g_2(\alpha\mid data)\propto \frac{1}{ (d+\sum_{i=1}^r {x_{i}^{\alpha}})^{r+c}} \alpha^{a+r-1} e^{-b\alpha} \prod_{i=1}^r {x_{i}^{\alpha+1}}.
$$
The logarithm of $g_2(\alpha|data)$ without the additive constant is

$$
\ln g_2(\alpha|data)=
-(r+c)\ln\left(d+\sum_{i=1}^r
x_i^\alpha \right)+(a+r-1)\ln(\alpha)-b\alpha+(\alpha+1)\sum_{i=1}^r \ln(x_i)\nonumber,
$$

Using Lemma 1 of Kundu (2007), it follows that
$$
\frac{d}{d\alpha^2}\ln\left(\sum_{i=1}^r
x_i^\alpha +d\right)\geq 0.
$$
Therefore the result follows.

\end{document}